\documentclass[twocolumn]{aastex631}
\usepackage[T1]{fontenc}
\usepackage{hyperref}
\usepackage{amsmath}
\usepackage{amssymb}
\usepackage{amsfonts}
\usepackage{hyperref}
	\hypersetup{
    unicode=true,          % non-Latin characters in Acrobat’s bookmarks
    pdftoolbar=true,        % show Acrobat’s toolbar?
    pdfmenubar=true,        % show Acrobat’s menu?
    pdffitwindow=false,     % window fit to page when opened
    pdfstartview={FitH},    % fits the width of the page to the window
    pdftitle={My title},    % title
    pdfauthor={Author},     % author
    pdfsubject={Subject},   % subject of the document
    pdfcreator={Creator},   % creator of the document
    pdfproducer={Producer}, % producer of the document
    pdfkeywords={keyword1, key2, key3}, % list of keywords
    pdfnewwindow=true,      % links in new PDF window
    colorlinks=true,        % false: boxed links; true: colored links
    linkcolor=MidnightBlue,     % color of internal links (change box color with linkbordercolor)
    citecolor=MidnightBlue,     % color of links to bibliography
    filecolor=MidnightBlue,     % color of file links
    urlcolor=MidnightBlue       % color of external links
}

\newcommand{\kms}{$\rm km s^{-1}$}

\newcommand{\HI}{\mbox{H\,{\sc i}}}

\newcommand{\OII}{\mbox{[O\,{\sc ii}]}}
\newcommand{\OIII}{\mbox{[O\,{\sc iii}]}}

\newcommand{\CIV}{\mbox{C\,{\sc iv}}}

\newcommand{\MgII}{\mbox{Mg\,{\sc ii}}}

\newcommand{\FeII}{\mbox{Fe\,{\sc ii}}}

\newcommand{\gotoqs}{GOTOQs}

\begin{document}
%\invertbackgroundtext
\title{H\textsc{i} 21-cm Absorption Associated with Foreground Galaxies on Top of Quasars}

\author{Labanya K. Guha}
\affiliation{Indian Institute of Astrophysics (IIA), Koramangala, Bangalore 560034, India}

\author{Raghunathan Srianand}
\affiliation{IUCAA, Postbag 4, Ganeshkhind, Pune 411007, Maharashtra, India}

\author{Rajeshwari Dutta}
\affiliation{IUCAA, Postbag 4, Ganeshkhind, Pune 411007, Maharashtra, India}

\correspondingauthor{Labanya K. Guha}
\email{labanya.guha@iiap.res.in}




\begin{abstract}
A systematic search for \HI\ 21-cm absorption in Quasar-Galaxy Pairs (QGPs) provides a powerful means to map the distribution of cold gas around high-redshift star-forming galaxies. Fiber spectroscopy of high-redshift quasars enables the serendipitous detection of foreground star-forming galaxies at extremely small impact parameters, forming a unique subset of QGPs known as Galaxies On Top Of Quasars (GOTOQs).  In this study, we present results from a pilot upgraded Giant Metrewave Radio Telescope (uGMRT) survey of three GOTOQs, where we achieved a remarkable 100\% detection rate of \HI\ 21-cm absorption. By combining our findings with existing literature, we establish that GOTOQs constitute a distinct population in terms of \HI\ 21-cm absorption, with significantly higher detection rates than those observed in Damped Lyman-$\alpha$ (DLA)-based or metal absorption-based searches. For the GOTOQs, we find a strong correlation between the line-of-sight reddening and the \HI\ 21-cm optical depth, characterized by \(\int \tau\, dv\, (\rm{km\,s^{-1}}) = 13.58^{+2.75}_{-2.35} E(B-V) + 0.68^{+1.06}_{-1.27}\), consistent with the Milky Way sightlines. We also show that the HI 21-cm optical depth declines with the impact parameter, and find a tentative trend for the HI 21-cm detection rates to also decline with the impact parameter. With upcoming wide-field spectroscopic surveys expected to substantially expand the catalog of known GOTOQs, the success of this pilot survey lays the foundation for constructing a statistically significant sample of intervening \HI\ 21-cm absorbers.
\end{abstract}



\section{Introduction}
\label{sec:intro}
Galaxy evolution is believed to be governed by the slowly varying equilibrium between the gas inflows from the intergalactic medium (IGM), gas outflows from the galaxy to the circumgalactic medium (CGM) and the IGM, and in situ star formation occurring within the galaxy \citep{Erb2008, Kennicutt2012, tumlinson2013,Peroux2020a}. Obtaining direct constraints on the gas inflow and outflow rates and how they evolve with redshift (i.e., cosmic time) is crucial for our understanding of galaxy evolution. In particular, \HI\ 21-cm absorption detected towards radio-loud sources are useful in probing the properties of cold neutral gas in and around galaxies and their redshift evolution.

The \HI\ 21-cm absorption serves as a powerful diagnostic tool for probing various physical properties of the neutral gas. It provides key insights into (1) the parsec-scale structure of the \HI\ gas through Very Long Baseline Interferometry (VLBI) observations \citep{srianand2013}; (2) the thermal state of the \HI\ gas \citep{heiles2003}  as the spin temperature ($T_s$) of neutral hydrogen, which governs the relative population of the hyperfine levels responsible for the 21-cm line, is influenced by the kinetic temperature ($T_k$) of the gas through collisional interactions and resonant scattering of Lyman-$\alpha$ photons\citep{Field1959, Liszt2001, Roy2006}; (3) the presence and strength of magnetic fields in the interstellar medium (ISM) via Zeeman splitting \citep{Heiles2004}; and (4) the filling factor of cold gas in both the ISM and the CGM.

Most of the detections of \HI\ 21-cm absorption at $z<1.5$ till date are based on searches towards strong Mg~{\sc ii} absorbers \citep[e.g.][]{briggs1983, Kanekar2009, gupta2009, Dutta2017MgII}. Based on these studies, the average detection rate of \HI\ 21-cm absorption in samples of strong \MgII\ systems (defined as rest equivalent width of the \MgII\ $\lambda2796$ line, $\rm W_{2796}\geqslant 1$\AA) is found to be $\sim 10-20$\% for an optical depth sensitivity of $\int \tau dv = 0.3$ \kms. This incidence is found to remain constant, within the measurement uncertainties, over the redshift range $z \simeq $ 0.3 - 1.5 \citep{Dutta2017MgII}.  Further, the \HI\ 21-cm absorption detection rate is shown to increase (i) with the equivalent width of \MgII\ and \FeII\ \citep{Dutta2017FeII}, (ii) with higher cut-offs in the equivalent width ratio of Mg~{\sc i}/\MgII\ and \MgII/Fe~{\sc ii} \citep{gupta2012}, and (iii) when the background quasar is dust reddened \citep{Carilli1998,Dutta2020redqso}. Typically, associated galaxies are found within $\approx$30 kpc of the lines of sight  \citep[see, for example, the compilation of][]{curran2016, Dutta2017}.  However, due to various biases associated with the optical pre-selection and lack of independent \HI\ column density measurements in most cases, interpreting the detection rate of \HI\ 21-cm absorption from such an ``absorber-centric'' sample is not straightforward.

To effectively map the distribution and the redshift evolution of the \HI\ 21-cm absorbers, conducting a blind survey of \HI\ 21-cm absorbers is thus essential. This has recently become possible thanks to the advent of sensitive, wide-band receivers on radio interferometers. \citet{Gupta2021} carried out a blind survey (covering a redshift path length of $\sim$13 at $z\sim$0.18) with the upgraded Giant Metrewave Radio Telescope \citep[uGMRT;][]{Gupta2017ugmrt}, and placed a constraint on the number of \HI\ 21-cm absorbers per unit redshift of $\leqslant$0.14 for $\int \tau dv\geqslant0.3$ \kms, corresponding to \HI\ column density, $\rm N_{H I} \gtrsim 5\times10^{19}\, cm^{-2}$ for $\rm{T}_s = 100$ K. Using the position of nearby photometric galaxies, they also concluded that the covering factor of \HI\ gas is $\leqslant$ 0.022 for the impact parameters (D) \footnote{Impact parameter (D) is the projected separation between the centre of the galaxy and background line of sight.} in the range 50 kpc $\leqslant$ D $\leqslant$ 150 kpc. \citet{Sadler2020} have detected 4 \HI\ 21-cm absorbers in their blind survey using Australian Square Kilometer Array Pathfinder \citep[ASKAP;][]{Hotan2021} covering the redshift path length of 21.37 (at a median $z\sim0.6$) for detecting typical Damped Lyman-$\alpha$ absorbers (DLAs) with spin temperature of 100 K. This provided the number of absorbers per unit redshift path of $0.19^{+0.15}_{-0.09}$. In the future, full survey data releases from ongoing blind surveys with ASKAP \citep{Allison2022} and MeerKAT\citep{Gupta2015MALS} will provide unbiased statistics on the redshift distribution of \HI\ 21-cm absorbers at $z\leqslant1.5$. 

However, it is important to keep in mind that purely based on geometric considerations, blind surveys are likely to yield a limited number of background sightlines that pass through the halos of foreground galaxies at low impact parameters. Therefore, complementary to absorber-centric searches, several studies have used close projected pairs of background quasars and foreground galaxies, known as quasar-galaxy pairs (QGPs), to map the distribution of cold atomic gas around low redshift ($z<0.4$) galaxies \citep{carilli1992, gupta2010, borthakur2011, zwaan2015, reeves2016, Dutta2017}. However, \HI\ 21-cm emission studies have shown that high-column-density \HI\ gas ($N\rm_{ H\,I} \gtrsim 10^{20}\, cm^{-2}$) is typically confined to within $\approx 2-3$ times the optical radius of isolated galaxies \citep{Irwin1995, zwaan2005, reeves2015}. Consequently, targeted “galaxy-centric” surveys focusing on galaxies at low impact parameters (i.e., within a few optical radii) are essential for probing the distribution of cold gas. Furthermore, observations indicate that the detection rate of \HI\ 21-cm absorption is highest at such low impact parameters, with the covering fraction of cold neutral gas reaching $\approx$24\% at D $\leqslant$15 kpc \citep{Dutta2017}, implying the importance of these focused surveys.

Here we design an experiment to probe the cold \HI\ gas close to galaxies using fiber-fed spectroscopic observation, which is expected to capture photons from all objects present within the projected area of the fiber in addition to the photons from the primary target. For example, the Sloan Digital Sky Survey \citep[SDSS;][]{York2000} spectroscopy of a distant quasar can detect nebular emission lines from foreground galaxies typically within D $\sim$ 2 kpc at redshift $z \sim 0.1$ and $\sim$10 kpc at $z \sim 1$. \citet{york2012} named such galaxies as Galaxies On Top Of Quasars \citep[GOTOQs; see also][]{Noterdaeme2010, straka2013, Straka_2015, Joshi2017, Joshi2018, Rubin2022, Das2024}. GOTOQs thus provide excellent targets to probe the cold \HI\ gas at low impact parameters to star-forming galaxies over a wide range of redshifts. This forms the main motivation of this pilot survey.

This paper is organized as follows. In section~\ref{sec:sample} we provide the basic details of our sample. Details of our uGMRT observations and data reduction are provided in section~\ref{sec:observations}. Analaysis and results are summarized in section~\ref{sec:results}. Section~\ref{sec:summary} discusses and summarizes the main results of our paper. Throughout this work we use flat $\Lambda$CDM cosmology with $\Omega_\Lambda = 0.7$, $\Omega_m = 0.3$, and $H_0 = 70 \,\rm{km\, s^{-1}\, Mpc^{-1}}$.

\section{Sample} \label{sec:sample}
The \gotoqs\ searched for \HI\ 21-cm absorption in this work are drawn from a  parent sample constructed by Guha et al. (in preparation). This parent sample, the largest of its kind till date, consists of about 500 \gotoqs\ at $z_{\text{\rm{GOTOQ}}} < 0.4$ from SDSS-DR16 \citep{alam2015}. Briefly, \gotoqs\ are identified from quasar spectra using the following procedure: the continuum is first subtracted using \texttt{QSmooth} \citep{Durovcikova2020}, and the regions around prominent quasar emission lines are masked. The continuum subtracted spectrum is then smoothed over three pixels to search for  possible foreground galaxy emission lines. A peak-finding algorithm detects the emission features, which are then matched to appropriately redshifted known nebular emission lines. Spectra with at least three matching lines detected with a significance of at least $3\sigma$ are marked as potential candidate \gotoqs\ and are finally visually inspected to confirm them as GOTOQs.

To identify associated \HI\ 21-cm absorption in \gotoqs, we cross-matched the \gotoqs\ with the Faint Images of the Radio Sky at Twenty-Centimeters (FIRST) 1.4 GHz \citep{white1997} and Very Large Array Sky Survey (VLASS) 3 GHz \citep{Gordon2021} radio source catalogs. We selected sources with a peak flux density exceeding 30 mJy and ensured that the expected \HI\ 21-cm signal was free from contamination by known strong Radio Frequency Interference (RFI) at uGMRT. The cross-matching criterion required the optical and radio centroids of the quasars to be within 1 arcsecond of each other. This process identified seven \gotoqs, one of which has been previously observed in the literature \citep{borthakur2010}. We are conducting a systematic search for cold gas in the remaining systems, with observations of four quasars completed using uGMRT in cycle 46. Details of these observed \gotoqs\ are provided in Table \ref{tab:observation_log}. The observations of one of these four systems (i.e., J0055+1408) are severely affected by scintillation, resulting in data that are inadequate for analysis. Consequently, this system is excluded from the study and we focus on presenting observations for the remaining three systems. 

\begin{figure*}
    \centering
    \includegraphics[width=\linewidth]{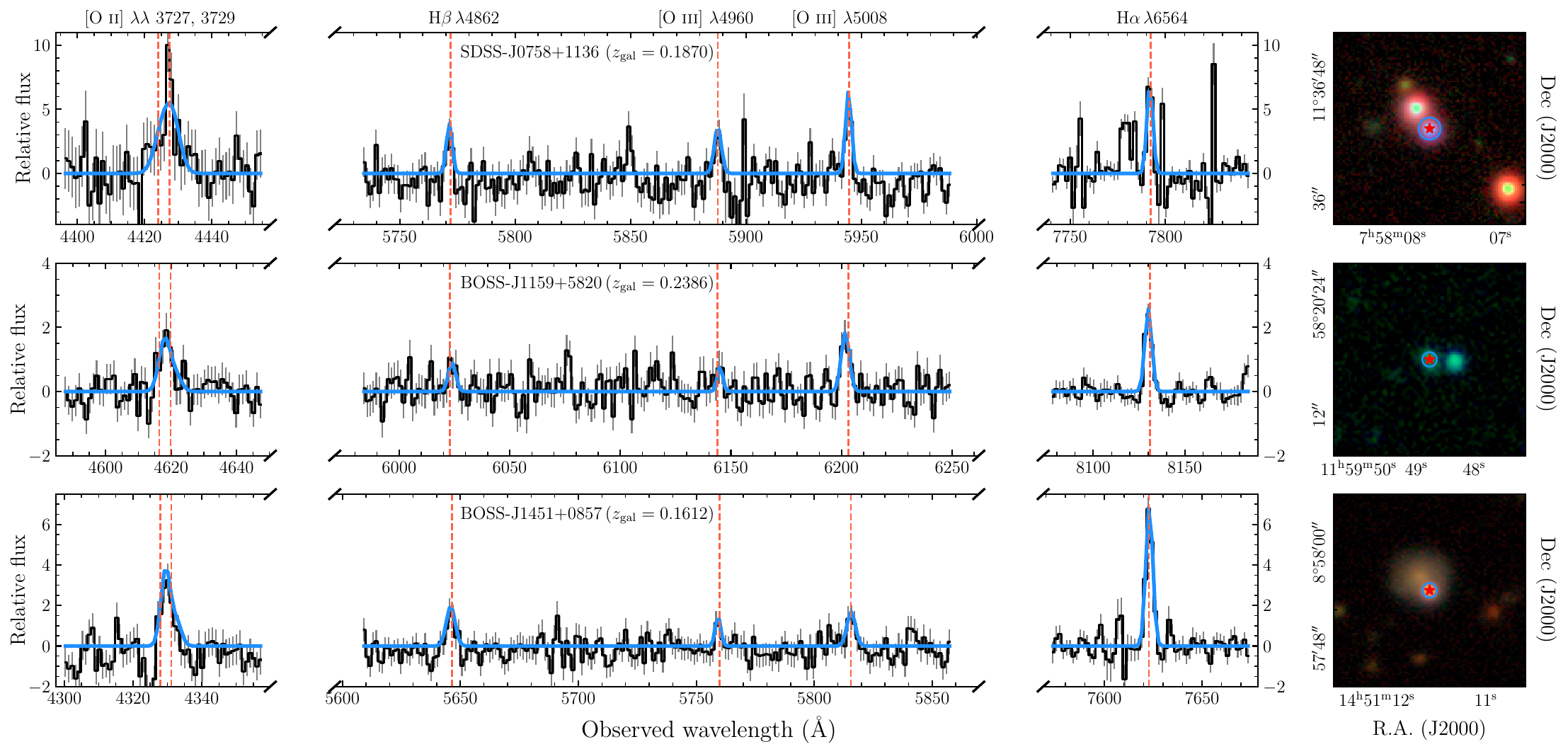}
    \caption{Optical properties of \gotoqs\ observed with the uGMRT. Each horizontal panel consists of two plots: the left plot displays the nebular emissions detected in the continuum-subtracted SDSS/eBOSS quasar spectrum where the prominent nebular emission lines from the foreground galaxy are marked by red dashed vertical lines and the fit to the data are shown in blue. Right plot shows a composite $griz$ band DESI-LIS image of the quasar field, with the quasar marked by a red star and the observing fiber outlined by a blue circle. Flux scales in the left panels are in the units of $\rm{ 10^{-17}\, ergs\, s^{-1}\, cm^{-2}\, \text{\AA}^{-1}}$.}
    \label{fig:21cm_gotoqs_sample}
\end{figure*}

Available optical data for these three \gotoqs\ are shown in Figure \ref{fig:21cm_gotoqs_sample}. Each horizontal panel in this figure consists of two plots: the left plot displays the nebular emission lines detected in the continuum-subtracted SDSS quasar spectrum. The prominent nebular emission lines from the foreground galaxy are marked by red dashed vertical lines and the fit to the data are shown in blue. The \OII\ $\lambda\lambda\, 3727, 3729$ emission doublets are fitted with double Gaussians with both the lines having same redshift and velocity widths, while other nebular emission lines are fitted using single Gaussian. The plot on the right shows a composite $griz$ band Dark Energy Survey Instrument - Legacy Imaging Survey \citep[DESI-LIS;][]{Dey2019DESI} image of the quasar field, with the quasar marked by a red star and the area covered by the observing fiber outlined by a blue circle.  The foreground galaxy is clearly visible in only one case (i.e., J1451+0857), where we could measure the impact parameter. In other two cases we could get only upper limit on the impact parameter based on the SDSS or Baryon Oscillation Spectroscopic Survey (BOSS) fiber sizes.

\begin{table*}
    \centering
    \begin{tabular}{lccccccr}
    \hline
    \hline
    (1) & (2) & (3) & (4) & (5) & (6) & (7) & (8)\\
    GOTOQ & $z_{\text{\rm{QSO}}}$ & $z_{\text{\rm{gal}}}$ & Observation date & On-source Time & $\delta v_{\rm ch}$ & RMS Noise & $f_{\rm peak}$\\
    &   &   &   & (mins) & ($\rm km\, s^{-1}$) & (mJy/beam) & (mJy/beam)\\
    \hline
    J005503.52+140806.50$^\star$ & 1.666 & 0.2015 & 30 November, 2024 & 111 & -- & -- & -- \\
    J075807.65+113646.00 & 0.569 & 0.1870 & 15 November, 2024 & 64 & 0.76 & 5.4 & 420 \\
    J115948.76+582020.00 & 1.280 & 0.2386 & 19 July, 2024 & 12 & 0.80 & 13.1 & 1389 \\
    J145111.51+085757.80 & 1.355 & 0.1612 & 19 July, 2024 & 234 & 0.75 & 3.0 & 36 \\
    \hline
    \end{tabular}
    \caption{Log of our uGMRT observations. Columns (1), (2), and (3) provide the \gotoqs\ system name, redshift of the background quasars, and the redshift of the foreground galaxies, respectively. Columns (4) and (5) provide the date of observations and the on-source observing time for the targets, respectively. Columns (6) and (7), respectively, indicate the channel width in the velocity space and the root-mean-square (RMS) noise of the spectrum. Column (8) lists the peak radio flux density of the background quasar in the uGMRT image. Observations of the system marked with $\star$, J0055+1408, are severely affected by scintillations.}
    \label{tab:observation_log}
\end{table*}

\begin{table*}[]
    \centering
    \caption{Properties of the GOTOQs analyzed in this work, along with those previously studied in the literature, are summarized as follows: Column (1) lists the coordinates of the background quasars, while Columns (2) and (3) provide the impact parameters of the foreground galaxies and their emission redshifts, respectively. Column (4) presents the integrated \HI\ 21-cm optical depth towards the background quasars associated with these galaxies. Column (5) represents the 90\% of the velocity width of the absorbing gas ($\Delta v_{90}$). Column (6) lists the gas phase metallicity of the foreground galaxies based on the nebular line ratios derived from the SDSS spectra.  Column (7) indicates the line-of-sight reddening of the quasars caused by the foreground galaxies, and Column (8) lists the references for the previously observed GOTOQs compiled in this study.}
    \begin{tabular}{cccccccccc}
    \hline
    \hline
    QSO & D(kpc) &$z_{gal}$ &$\int \tau dv$ (km/s) & $\Delta v_{90}$ (km/s) & $12 + \log{(O/H)}$ & ${\rm E(B-V)}$ & Ref\\
    (1) & (2) & (3) & (4) & (5) & (6) & (7) & (8) \\
    \hline
     J075807.65+113646.0 & $<4.7$ & 0.1870 & $1.64\pm0.07$ & 27.3 & $8.32$ & $-0.071\pm0.002$ & This work\\
     J115948.76+582020.0 & $<3.8$ & 0.2386 & $7.39\pm0.04$ & 11.9 & $8.52$ & $+0.353\pm0.007$ & This work\\
     J145111.51+085757.8 & 8.0    & 0.1612 & $2.54\pm0.28$ & 8.4  & $8.70$ & $+0.252\pm0.007$ & This work  \\
     J011322.69+251853.3 & 6.0    & 0.2546 & 11.05         & .... & ....   & $+0.597\pm0.002$ & \citet{Hu2024Intervening21cm} \\ % Hu+2024
     J104257.58+074850.5 & 1.7    & 0.0332 & 0.19          & 4.0  & $8.42$ & $+0.142\pm0.004$ & \citet{borthakur2010} \\ %Borthakur+2010
     J124157.54+633241.6 & 11.1   & 0.1430 & 2.90$\pm$0.16 & 52.8 & $8.7$  & $+0.195\pm0.002$ & \citet{gupta2010} \\ %Gupta+2010
     J130028.53+283010.1 & 9.4    & 0.2229 & $\leqslant0.239$    & .... & ....   & $-0.089\pm0.003$ & \citet{Dutta2017} \\ % Dutta+2017 
     J143806.79+175805.4 & 7.5    & 0.1468 & 4.89$\pm$0.19 & 19.7 & ....   & $+0.060\pm0.003$ & \citet{Dutta2017} \\ % Dutta+2017 
     J144304.53+021419.3 & $<5.1$ & 0.3714 & 3.40$\pm$0.10 & 16.5 & $8.4$  & $+0.433\pm0.002$ & \citet{gupta2013} \\ %Gupta+2013
     J163956.35+112758.7 & 5.1    & 0.0790 &15.70$\pm$0.13 & 28.0 & $8.47$ & $+1.112\pm0.004$ & \citet{srianand2013} \\ %Srianand+2013
    \hline
    \end{tabular}
    \label{tab:summary}
\end{table*}

\section{Observations and Data Reduction} \label{sec:observations}

The sample was observed using Band 5 of uGMRT between July and December 2024 (Proposal ID: 46\_058). The sources were observed for $\approx1-6$ h depending on their radio continuum flux density to reach a similar optical depth sensitivity (see Table~\ref{tab:observation_log}). Standard flux density and gain calibrators were observed in regular intervals during the course of the observations. The data were obtained in two parallel hand correlations. The spectral setup used a bandwidth of 6.25 MHz split into 2048 channels giving a channel width of 3 kHz. The data reduction was carried out using the Common Astronomy Software Package \citep[{\sc CASA}, version 6.6;][]{McMullin2007}, and involved standard procedures of data editing, gain and bandpass calibration, and iterative cycles of self-calibration and imaging. In addition to manual data editing in {\sc CASA}, the {\sc AOFLAGGER} package \citep{Offringa2012} was used to flag data affected by RFI. After subtracting the continuum emission from the self-calibrated visibilities, the spectral cubes were obtained using Briggs Robust=0 weighting. The typical angular resolution of the cubes and continuum images for this weighting scheme is $\approx2-3$ arcsec.  The radio contours overlayed on the optical images are shown in Figure \ref{fig:radio_contours}. The H\,{\sc i} 21-cm absorption spectra were extracted towards the peak of the continuum flux density of the radio sources. If required a low-order polynomial was fit to the line-free channels of the spectra to remove any residual continuum emission.

\section{Analysis and results}
\label{sec:results}

\begin{figure*}[htbp]
    \centering
    % First figure
    \begin{minipage}[b]{0.33\textwidth}
        \centering
        \includegraphics[width=\textwidth]{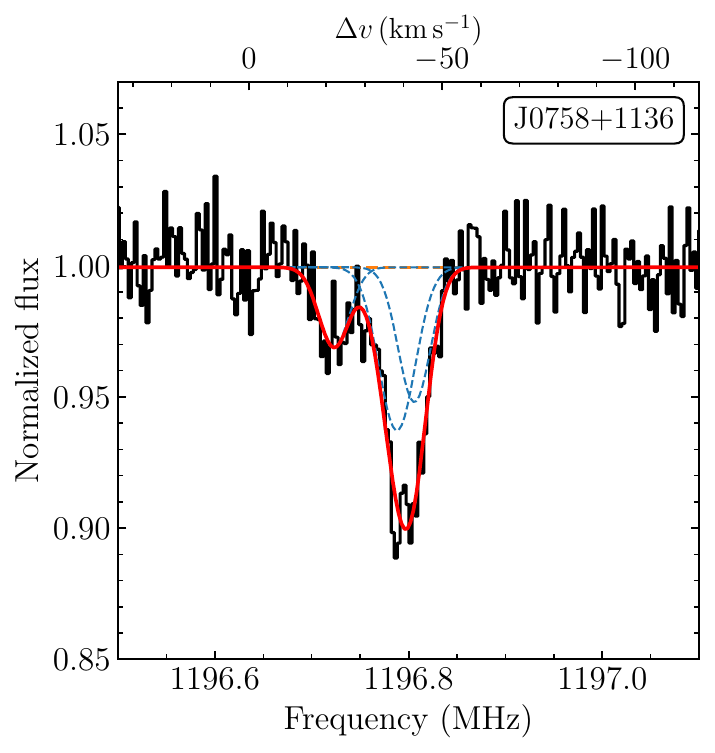}
    \end{minipage}
    \hfill
    % Second figure
    \begin{minipage}[b]{0.32\textwidth}
        \centering
        \includegraphics[width=\textwidth]{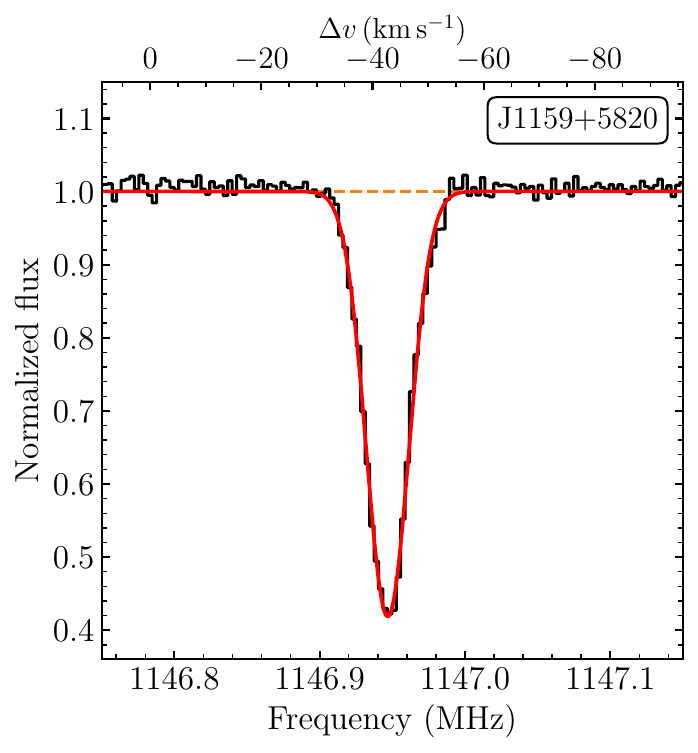}
    \end{minipage}
    \hfill
    % Third figure
    \begin{minipage}[b]{0.32\textwidth}
        \centering
        \includegraphics[width=\textwidth]{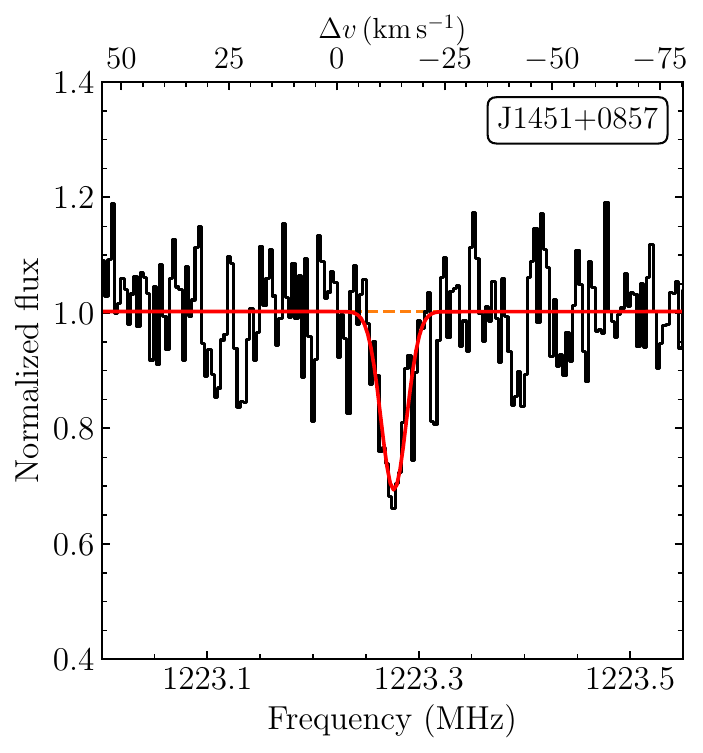}
    \end{minipage}
    \caption{ \HI\ 21-cm absorption spectra of three \gotoqs\ : J0758+1136 (left panel), J1159+5820 (middle panel), and J1451+0857 (right panel). The individual Gaussian components are shown as blue dashed lines, while the overall fits to the absorption profiles are shown in solid red lines. The upper $x$-axis indicates the corresponding velocities relative to the emission redshifts of the foreground galaxies.} 
    \label{fig:21cm_absorption}
\end{figure*}

We detect \HI\ 21-cm absorption toward the background radio-loud quasars associated with all three GOTOQs for which we could secure reliable spectra with uGMRT. Figure \ref{fig:21cm_absorption} provides the \HI\ 21-cm absorption spectra for these GOTOQs: J0758+1136 (left panel), J1159+5820 (middle panel), and J1451+0857 (right panel). The continuum-normalized observed spectra are plotted in black, while the overall multiple component Gaussian fits to the absorption profiles are shown in red. The individual Gaussian components are shown using blue dashed lines. The full width at half maximum (FWHM) of individual Gaussian components ranges from ${7.0\pm1.3}$ km s\(^{-1}\) to ${11.1\pm2.0}$ km s\(^{-1}\), with a median value of 9.3 km s\(^{-1}\). The width of absorption lines can serve as a constraint on the temperature of the absorbing gas, assuming that thermal motions are the primary driver of line broadening. However, this estimate represents an upper limit, as additional factors such as turbulence can also contribute to the observed broadening. Assuming a representative \(\Delta v_{\text{FWHM}} = 10\) km s\(^{-1}\) for GOTOQs, we derive an upper limit on the kinetic temperature of the gas as \(T_k \leqslant 2185\) K \citep{borthakur2010}. This is higher than the $T_s$ constraint we have obtained using the correlation between $E(B-V)$ and $\int \tau dv$ (see section 4.3). This implies either the presence of non-thermal motions and/or hidden narrow components, which can be resolved by obtaining spectra at higher resolution and signal-to-noise ratio. 

The integrated \HI\ 21-cm optical depths are provided in Column (4) of Table \ref{tab:summary}. The observed integrated \HI\ 21-cm optial depths are consistent with these sightlines being damped Lyman-$\alpha$ (DLA; i.e., $\rm \log{[N_{ H\,I} / cm^{-2}]} \geqslant 20.3$) absorbers assuming a spin temperature of the gas, $T_s = 100$ K, and a covering factor (i.e., \( f_c\)\footnote{Covering factor is usually defined as the fraction of the background radio source occulted by the absorbing gas.})  of unity. This is in line with the finding of \citet{kulkarni2022} that \gotoqs\ produce DLA (or sub-DLA, i.e., $\rm 19 \leqslant \log{[N_{ H\,I} / cm^{-2}]} < 20.3$) absorption in the ultraviolet spectra of the background quasars. The properties of the GOTOQs analyzed in this work and those compiled from the literature, are summarized in Table \ref{tab:summary}.

Nine of the ten GOTOQs show detection of \HI\ 21-cm absorption. The weakest \HI\ 21-cm absorption was observed towards J1042+0748  with a measured $\int \tau dv$ of 0.19 \kms \citep{borthakur2010}. If we consider the limiting $\int \tau dv$ of 0.3 \kms\ as typically considered in the literature \citep[e.g.,][]{Dutta2017}, then the \HI\ 21-cm detection rate is $\sim$80\% for GOTOQs.  This is by far the largest detection rate for \HI\ 21-cm absorption in any population of isolated, intervening galaxies. Such large detection rates (i.e., $\sim$ 80\%) are also seen  towards galaxy mergers at similar redshifts \citep[]{Dutta2018, Dutta2019}. The optical depth measured in the case of detections are consistent with them being DLAs (assuming $f_c = 1$ and $T_s = 100$ K), except for J1042+0748, where VLBI observations constrain the \HI\ column density to $\rm N_{H\, I} \leqslant 1.5\times10^{20}\, cm^{-2}$ \citep{borthakur2010}.

To characterize the kinematic spread of the \HI\ 21-cm absorbing gas, we estimate \(\Delta v_{90}\), defined as the velocity range within which 90\% of the total integrated optical depth is contained. The computed \(\Delta v_{90}\) values are listed in Column (5) of Table \ref{tab:summary}. For the GOTOQs, \(\Delta v_{90}\) ranges from 4 km s\(^{-1}\) to 52.8 km s\(^{-1}\), with a median value of 18.1 km s\(^{-1}\). This is consistent with the distribution of low-redshift (\(z \leqslant 0.4\)) \HI\ 21-cm absorbers associated with DLAs, strong \FeII /\MgII\ metal absorbers, and quasar-galaxy pairs. In the sample compiled by \citet{Dutta2017FeII}, \(\Delta v_{90}\) ranges from 10 km s\(^{-1}\) to 59 km s\(^{-1}\) (median 11 km s\(^{-1}\)) for low-\(z\) DLAs, from 4.5 km s\(^{-1}\) to 39 km s\(^{-1}\) (median 19.1 km s\(^{-1}\)) for low-\(z\) strong \FeII /\MgII\ absorbers, and from 1.1 km s\(^{-1}\) to 76 km s\(^{-1}\) (median 19.7 km s\(^{-1}\)) for low-\(z\) quasar-galaxy pairs. This is expected as \gotoqs\ are usually strong \FeII /\MgII\ absorbers \citep{Guha2022b, Guha2024b} and are typically associated with DLAs or sub-DLAs \citep{kulkarni2022}. However, note that, compared to the high-redshift ($z>0.4$) \HI\ 21-cm absorbers associated with \FeII/\MgII\ absorption or DLAs, low-redshift \HI\ 21-cm absorbers have significantly less spread in the velocity space \citep{Dutta2017FeII}. 

\subsection{Integrated optical depth versus impact parameter}
In this subsection, we examine the dependence of the integrated \HI\ 21-cm optical depth around galaxies on the impact parameter. Among the three new detections reported here, the impact parameter is precisely measured only for J1451+0857, where the foreground galaxy is clearly visible in the DESI-LIS images. In the other two cases, the foreground galaxies are not detected in the DESI-LIS images, likely because they are at a very low impact parameter and outshined by the bright background quasars. Precise measurements of the impact parameters in such cases typically rely on triangulation methods \citep{Noterdaeme2010, Guha2024b}. However, for now, we provide an upper limit on the impact parameters based on the fiber size, as the fiber captures nebular emission from the foreground galaxy, thereby constraining the maximum possible extent of the impact parameters. This is also the case for the GOTOQ J1443+0214 in the literature sample. To better investigate the dependence of the integrated optical depth on the impact parameter, we incorporate additional quasar-galaxy pairs with $D \leqslant 40$ kpc previously studied by \citet{Dutta2017}. We consider only quasar-galaxy pairs where the redshift of the background quasar is spectroscopically confirmed.

\begin{figure}
    \centering
    \includegraphics[width=\linewidth]{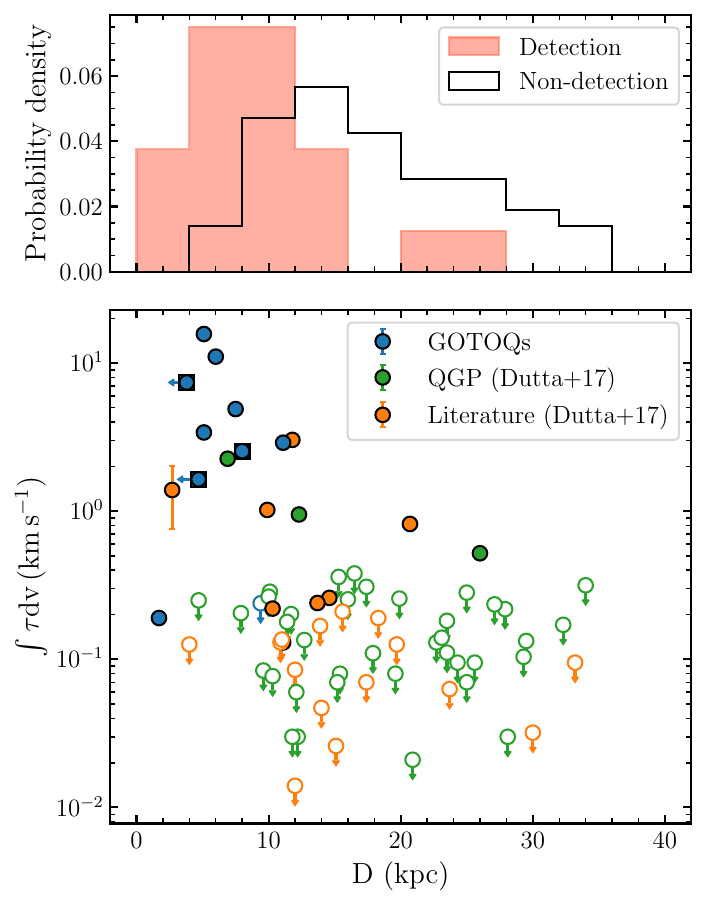}
    \caption{{\it Lower panel}: Integrated \HI\ 21-cm optical depth as a function of the impact parameter of galaxies. Blue points represent the GOTOQs, with newly observed GOTOQs using uGMRT highlighted by square boxes, while the remaining data points are taken from the literature. Green points indicate the quasar-galaxy pairs studied by \citet{Dutta2017}, and orange points correspond to quasar-galaxy pairs from the literature and compiled in \citet{Dutta2017}. Filled points denote \HI\ 21-cm detections in absorption, whereas hollow points represent the $3\sigma$ upper limits on the integrated optical depth. {\it Upper panel}: Normalized histogram of the impact parameter distribution for \HI\ 21-cm detections and non-detections.}
    \label{fig:taudv_vs_rho}
\end{figure}

In the lower panel of Figure \ref{fig:taudv_vs_rho}, we show the integrated \HI\ 21-cm optical depth as a function of the impact parameter of galaxies. Blue points represent the GOTOQs, with those newly observed using uGMRT highlighted by square boxes. Green points indicate the quasar-galaxy pairs studied by \citet{Dutta2017}, and orange points correspond to quasar-galaxy pairs from the literature and compiled in \citet{Dutta2017}. Filled points denote \HI\ 21-cm detections in absorption, whereas hollow points represent the $3\sigma$ upper limits on the integrated optical depth. In the upper panel of Figure \ref{fig:taudv_vs_rho}, we show the normalized histogram of the impact parameter distribution for \HI\ 21-cm detections (filled pink) and non-detections (black outline). It is apparent from this figure that the \HI\ 21-cm absorption is detected predominantly in cases where the quasar line-of-sight passes through the immediate vicinity of the foreground galaxies. A Kendall-$\rm \tau$ rank correlation test including the upper limits \citep{Privon2020} between the impact parameter and the integrated \HI\ 21-cm optical depth yields a strong anti-correlation between the two with the rank correlation coefficient $\tau_r = -0.353$ and $p$-value of $\approx 9\times 10^{-6}$ with the null hypothesis that they are not correlated. By including new samples of GOTOQs that probe small impact parameters, we confirm the anti-correlation between HI 21-cm optical depth and impact parameter reported in previous studies \citep{curran2016, Dutta2017} at a much higher significance level.  A KS test comparing the impact parameter distributions of galaxies with \HI\ 21-cm detections to those without shows that the underlying distributions are statistically very different ($p$ value of $\approx 3\times10^{-4}$, with the null hypothesis that two distributions are identical).  Similarly, based on the KS test ($p$-value = 0.356), we find no significant redshift dependence in the detection of \HI\ 21-cm absorption over the redshift range $0 < z < 0.4$.

\subsection{Detection rate of cold neutral gas around galaxies}
\begin{figure}
    \centering
    \includegraphics[width=\linewidth]{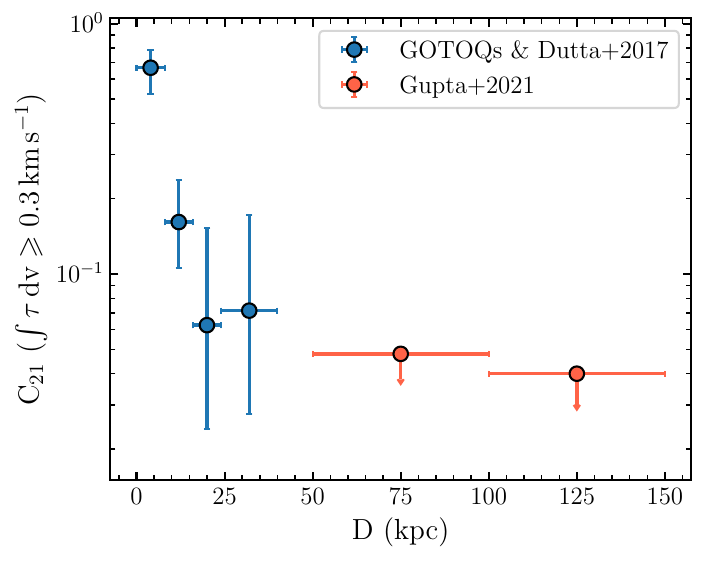}
    \caption{Covering fraction of \HI\ 21-cm absorbers ($\rm C_{21}$) as a function of impact parameter. Blue points show $\rm C_{21}$ across four impact parameter bins ($0-8$, $8-16$, $16-24$, and $24-40$ kpc) for integrated optical depth sensitivities of 0.3 $\rm km\,s^{-1}$ for the GOTOQs and the QGPs studied and compiled from literature by \citet{Dutta2017}. Red points are taken from \citet{Gupta2021}. }
    \label{fig:HI_fc}
\end{figure}

In this section, we examine the detection rate ($C_{21}$) of the cold \HI\ gas, as traced by \HI\ 21-cm absorption, around galaxies as a function of impact parameter (D). The detection rate, $C_{21}$, is defined as the fraction of systems with \HI\ 21-cm detections satisfying $\rm \int \tau dv \geqslant 0.3 \, km \, s^{-1}$. 
Figure \ref{fig:HI_fc} shows $C_{21}$ as a function of impact parameter bins ($0-8$, $8-16$, $16-24$, and $24-40$ kpc) for the GOTOQs and QGPs studied and compiled from the literature by \citet{Dutta2017}. Additionally, red points from \citet{Gupta2021} represent $C_{21}$ over larger impact parameter bins of $50-100$ and $100-150$ kpc.  The error bars on $C_{21}$ represent the $\rm 1\sigma$ Wilson binomial confidence intervals, while those on $D$ reflect the bin size. The figure clearly shows a steep decline in $C_{21}$ with increasing impact parameter as the measured $C_{21}$ values are $67^{+12}_{-14}\%$, $16^{+8}_{-6}\%$, $6^{+9}_{-4}\%$  and  $7^{+10}_{-4}\%$ for the impact parameter bins of $0-8$, $8-16$, $16-24$, and $24-40$ kpc. There is no \HI\ 21-cm absorption detected beyond this impact parameter and upper limits on $C_{21}$ of $\leqslant0.048$ and $\leqslant0.040$ over the impact parameter bins of $50-100$ kpc and $100-150$ kpc, respectively, were obtained by \citet{Gupta2021}. A Kendall $\tau$ rank-correlation test, including the upper limits \citep{Privon2020}, confirms an anti-correlation between $C_{21}$ and $D$, with a rank correlation coefficient of $\rm \tau_r = -0.814$ and a $p$ value of 0.021. To account for measurement uncertainties, we generated Monte Carlo realizations of the dataset. Symmetric errors on $D$ were sampled uniformly over the bin, while asymmetric errors on $C_{21}$ were represented using split-normal distributions. Upper limits in $C_{21}$ were sampled uniformly between zero and the measured value. For each realization, the Spearmann correlation coefficient and its $p$-value were computed. Repeating this process across 10000 realizations produced distributions of correlation coefficients and $p$-values that incorporate both symmetric and asymmetric measurement uncertainties, enabling robust estimation of the correlation and its uncertainty. This resulted in the rank correlation coefficient of $r_s = -0.886^{+0.229}_{-0.057}$ and the $p$ value of $0.019^{+0.092}_{-0.014}$.

Note, $C_{21}$ will depend upon the following factors: (i) covering factor of gas with $N$(\HI) sufficiently high to produce \HI\ 21-cm absorption, (ii) fraction of cold gas contributing to this $N$(\HI), and (iii) structure of the background radio source. Independent measurement of $N$(\HI) either through UV spectra along the quasar line-of-sight or through \HI\ 21-cm emission, along with higher spatial resolution VLBI observations to resolve the structure of background radio sources, would be highly valuable for interpreting the results.

\subsection{Line of sight reddening, integrated optical depth, and spin temperature}
\label{sec:redenning}

\begin{figure}
    \vskip0.01in
    \centering
    \includegraphics[width=\linewidth]{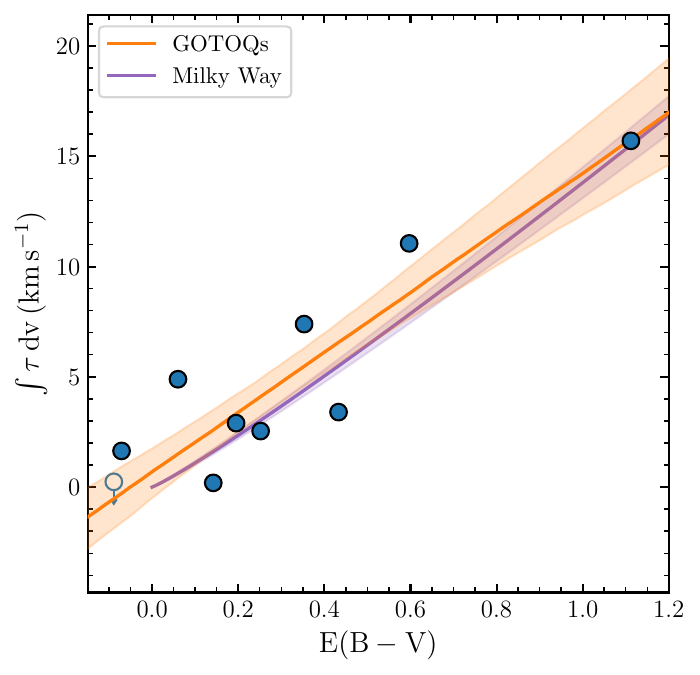}
    \caption{Integrated optical depth for \HI\ 21-cm absorption in GOTOQs compared to the color excess, \(E(B-V)\), caused by line-of-sight reddening. Occurrence of negative \(E(B-V)\) values indicate possible systematic uncertainty of $\sim0.1$ mag due to the quasar-to-quasar SED variations. However realistic estimation of error in E(B-V) will be achieved if we can use simultaneous SED measurements covering rest UV-to-IR range. The solid orange line represents the best-fit relation, while the shaded region indicates the \(1\sigma\) uncertainty in the fit. The purple line and the corresponding shaded region corresponds to the best-fit relation taken from \citet{Liszt2019} for the Milky Way \HI\ 21-cm absorbers.}
    \label{fig:taudv_vs_ebv}
\end{figure}

We estimate the line-of-sight reddening of the background quasar caused by foreground galaxies in the \gotoqs\ sample, assuming the gas follows an extinction curve similar to that of the Small Magellanic Cloud \citep[SMC;][]{Gordon_2003}. Note that for the rest frame wavelength range considered here, the choice of extinction curve has no significant effect on the below results. More robust measurements of reddening, including the actual extinction curve, can be obtained by including higher spatial resolution data that cover a wider range of wavelengths. To determine the line-of-sight reddening, we fitted the optical quasar spectra from SDSS using a standard quasar spectrum template \citep{Selsing2016}, redshifted to match the  redshift of the background quasar. The SMC extinction curve was then applied at the redshift of the foreground galaxies, with the V-band extinction coefficient ($A_V$) as the sole free parameter, alongside a multiplicative scaling factor, following the method described by \citet{srianand2008a}. Column (7) of Table \ref{tab:summary} provides the color excess, $E(B-V)$, for the background quasars resulting from line-of-sight reddening (assuming $R_V = 2.74$). The resultant fits for the GOTOQs studied in this paper are shown in Appendix \ref{sec:appendix_reddening}. For the GOTOQs, $E(B-V)$ ranges from $-0.089$ to $1.112$, with a median value of 0.223. Note the negative value of $E(B-V)$ in some cases is the consequence of intrinsic variance of the quasar spectral energy distribution (SED). Further note that among the GOTOQs analyzed in this study, only J1451+0857 exhibits strong \MgII\ absorption ($\rm W_{2796} = 0.72\pm0.10$\AA) at $z = 0.6633$. In the literature sample, J1241+6323 shows multiple \MgII\ and \CIV\ absorbers along the line of sight, while J1438+1758 and J1443+0214 exhibit \CIV\ absorption at $z = 1.7874$ and \MgII\ absorption at $z = 1.8024$, respectively. Given the difficulty in quantifying the individual contributions of these absorbers to the line-of-sight reddening, we assume that the GOTOQs are the dominant contributors, as the lines of sight intersect the disks of the foreground galaxies.

In Figure \ref{fig:taudv_vs_ebv}, we show the integrated optical depth of \HI\ 21-cm absorption for the \gotoqs\ as a function of color excess, $E(B-V)$. The figure shows a strong correlation between the integrated optical depth and line-of-sight reddening. A Kendall-$\tau$ rank correlation test between $E(B-V)$ and $\int \tau\, dv$ yields a positive correlation coefficient of $r_{\tau} = 0.669$ with the $p$ value of 0.007. We perform a linear fit between \(E(B-V)\) and \(\int \tau\, dv\) for the GOTOQs, incorporating upper limits, following the prescription described in \citet{Guha2022a} and \citet{Dutta2020}. We define the likelihood function as the product of likeliehoods for the detections and the non-detections, as follows:
\begin{equation*}
\begin{aligned}
    & \mathcal{L}(\Upsilon) = \left(\prod^{n}_{i=1} \frac{1}{\sqrt{2\pi \sigma_i^2}} \rm{exp}\left\{ {-\frac{1}{2}\left[\frac{\Upsilon_i - \Upsilon(\mathcal{E}_i)}{\sigma_i}\right]^2} \right\}  \right) \\
     & \times \left(\prod^{m}_{i=1} \int_{-\infty}^{\Upsilon_i} \frac{d\Upsilon^{\prime}}{\sqrt{2\pi \sigma_i^2}} \rm{exp} \left\{ -\frac{1}{2} \left[\frac{\Upsilon^{\prime} - \Upsilon(\mathcal{E}_i)}{\sigma_i}\right]^2\right\} \right)
\end{aligned}
\end{equation*}
with $\Upsilon_i = \left(\int \tau\, dv\right)_i$, and $\mathcal{E}_i = E(B-V)_i$, corresponding to the $i$-th measurement, whereas $\sigma_i = \sqrt{\sigma_{int}^2 + \sigma_{{\Upsilon_i}}^2}$ is defined so as to allow for the intrinsic scatter in the relation. The resulting best-fit relation is given by,
\begin{equation}
\int \tau\, dv\, (\rm{km\, s^{-1}}) = 13.58^{+2.75}_{-2.35} E(B-V) + 0.68^{+1.06}_{-1.27}
\end{equation}
with an intrinsic scatter of \(\sigma_{int} = 2.69^{+0.84}_{-0.66}\, \rm{km\, s^{-1}}\). The solid orange line represents the best-fit relation, while the shaded region denotes the \(1\sigma\) uncertainty. Interestingly, the best fit relationship obtained is very much identical to the relationship obtained for the galactic sightlines by \citet{Liszt2019}, \(\int \tau\, dv = (13.8\pm0.7)~E(B-V)^{1.10\pm0.03} \) (see purple line in Figure~\ref{fig:taudv_vs_ebv}), where \(E(B-V)\) is derived using far-IR measurements. This likely indicates that the physical conditions prevailing in GOTOQ absorbers may be similar to what we see in the local ISM. Notably, in absorber-centric surveys, while stacked spectrum of systems exhibiting \HI\ 21-cm absorption along their line-of-sight tend to be significantly redder than those without, the correlation between integrated \HI\ 21-cm optical depth and \(E(B-V)\) for individual cases is not found to be statistically significant \citep{Dutta2017FeII}.

Using the standard relationship between \(N(\text{H~{\sc i}})\) and \(\int \tau dv\), we obtain:  \(N(\text{H\,{\sc i}}) [\text{cm}^{-2}] \simeq 2.48^{+0.50}_{-0.43} \times 10^{19} \times \frac{T_s}{f_c} \times E(B-V) \). Here, \(T_s\) is the spin temperature in K, and \(f_c\) represents the gas covering factor of the background radio sources. For our discussion, we assume \(f_c = 1\). Previous studies by \citet{Gordon_2003} and \citet{Gudennavar2012} have established empirical relationships between \(N(\text{H~{\sc i}})\) and \(E(B-V)\) in the local universe. Specifically, the proportionality constants are \(3.05 \times 10^{21}\) , \(1.11 \times 10^{22}\), and \(3.61 \times 10^{22}\) for the Milky Way \citep{Gudennavar2012}, Large Magellanic Cloud (LMC), and SMC \citep{Gordon_2003}, respectively. Based on these relations, we estimate the harmonic mean spin temperature (\(T_s\)) for gas along GOTOQ sightlines to be approximately 123 K, 448 K, and 1456 K, assuming dust properties similar to those of the Milky Way, LMC, and SMC, respectively. Comparing these results with direct measurements of \(N(\text{H~{\sc i}})\) will not only provide further constraints on \(T_s\) but also help characterize the dust properties in GOTOQs. 

Given that gas-to-dust ratios in galaxies strongly depend on gas-phase metallicity \citep{Remy-Ruyer2014}, the metallicity of \gotoqs\ can provide additional constraints on the $T_s$ when compared to those of the SMC, LMC, and Milky Way. For the three \gotoqs\ in our sample, we estimate the gas-phase metallicity using nebular line diagnostics, incorporating \OII, \OIII, $\rm H\beta$, and $\rm H\alpha$ emission lines. This analysis is performed with the publicly available \textsc{python-izi} code \citep{Mingozzi2020}, utilizing the photoionization grid of \citet{Levesque2010}. Since [N \textsc{ii}] lines are not always detected in these cases, the derived metallicities exhibit a double-valued degeneracy. To ensure consistency, we adopt the upper-branch metallicities. The resulting values are listed in Column (6) of Table \ref{tab:summary}. The metallicities of the GOTOQs reported in the literature are obtained from their respective sources. The median gas-phase metallicity of the \gotoqs\ in our sample is \(12 + \log{(O/H)} = 8.47\), corresponding to \( \approx 0.6Z_\odot\). This value is comparable to the metallicity of the LMC \citep[\(12 + \log{(O/H)} = 8.5\);][]{Roman-Duval2022}, suggesting that the expected spin temperature in \gotoqs\ will be more closer to the value we obtained from the Milky Way or LMC relationship between $N$(\HI) and $E(B-V)$. However, we caution that the gas-phase metallicities derived from nebular emission lines in the star-forming region may not precisely match with the metallicites along the quasar line-of-sight probing the cold neutral medium, as assumed in this analysis. Nonetheless, since the quasar sightlines for the \gotoqs\ pass through the star-forming disks, any discrepancies are likely to be minimal.

Lastly, we also measured $E(B-V)$ using SED fitting for the GOTOQ lines-of-sight having $N$(\HI) measurements from UV spectroscopy obtained by \citet{kulkarni2022}. We do not find a significant correlation (Kendall $\tau$ rank correlation coefficient, $\tau_r = 0.25$ with $p$ value of 0.38) between $N$(\HI) and $E(B-V)$ for this sample. This lack of correlation could be because of a few reasons: (i) small sample size (i.e., 8 measurements); (ii) bias against reddened lines-of-sights in order to facilitate UV spectroscopy; and (iii) contribution to the \HI\ column density from neutral gas components other than that probed by the \HI\ 21-cm absorption, which is more sensitive to the colder neutral gas. It will be interesting to increase the number of $N$(\HI) measurements for \gotoqs\ using UV spectroscopy in the future to check whether a tight correlation exists between $N$(\HI) and $E(B-V)$. Finally, we point out that similar results were also found for high-$z$ C~{\sc i}-selected DLAs by \citet{Ledoux2015}.

\section{Discussions and conclusions}
\label{sec:summary}
In this study, we have presented the results of our uGMRT search for intervening \HI\ 21-cm absorption associated with \gotoqs. Our key results are summarized below.  

%$\bullet$
\vskip 0.1 in
\noindent{\bf Detection rate:} We report three new detections of intervening \HI\ 21-cm absorption at \( z < 0.4 \) associated with \gotoqs. This yields a 100\% detection rate. When combining our results with previously studied \gotoqs\ in the literature, we find that \HI\ 21-cm absorption is detected in 9 out of 10 \gotoqs\ at \( z < 0.4 \), corresponding to a detection rate of $\approx$90\% ($\approx$80\% if we use $\int\tau dv$ sensitivity limit of 0.3 \kms). On the other hand, if we consider all the QGPs within impact parameter of 8 kpc, we find the detection rate to be $\approx$67\%. Non-detections at low impact parameters are predominantly around low-surface brightness galaxies, where the \HI\ content and extent may be less. When it comes to absorption-centric searches, the detection rate in the case of $z<1$ DLAs (with impact parameters up to 45 kpc) is $\approx$61\% \citep[see Table 5 of][]{Dutta2017}. The detection rate of \HI\ 21-cm absorption from strong Mg~{\sc ii} absorbers is $\approx$18\%, where additional constraints on Fe~{\sc ii} equivalent width can make the detection rate go up to $\approx$50\% \citep{Dutta2017MgII, Dutta2017FeII}. Therefore, the highest detection rate of intervening \HI\ 21-cm absorption is in the sample of GOTOQs, which can be attributed to the low impact parameter probed around star-forming galaxies ensured by the selection method itself.

\vskip 0.1 in
 \noindent{\bf Impact parameter dependence of the optical depth \& detection rate:}
 By combining our data with the existing literature, we find that both the velocity-integrated optical depth and the detection rate of \HI\ 21-cm absorption decline rapidly with increasing impact parameter.  The low detection rate of \HI\ 21-cm absorption at high impact parameters could explain the low detection rate of \HI\ 21-cm absorption in ongoing large surveys using MeerKAT and ASKAP. On the other hand, in the case of low-z DLAs showing \HI\ 21-cm absorption, the associated galaxies are identified even up to 45 kpc. Note that DLA samples usually probe relatively higher redshifts compared to the QGPs. Therefore, differences in the impact parameter distribution could arise from (i) possible redshift dependencies, (ii) some of the DLAs originating from galaxies undergoing interactions that increase the \HI\ cross-section, and (iii) the identified nearby bright galaxies not being the actual host galaxies of the DLAs. Therefore, detailed host galaxy studies of absorber centric \HI\ 21-cm detections using integral field spectroscopic studies will provide important clues to understand the cross-section of cold \HI\ gas around galaxies. 

\vskip 0.1 in
\noindent{\bf Dust reddening and \HI\ 21-cm optical depth:}
 We identify a strong correlation between the line-of-sight reddening due to dust and the velocity-integrated optical depth of \HI\ 21-cm absorption in \gotoqs. Interestingly, the observed correlation is very similar to the known relationship of the Milk way. Confirming this trend with a larger sample of GOTOQs will be interesting. Further, it will be interesting to check whether these absorbers also follow a strong correlation between E(B-V) and $N$(\HI). However, in almost all the GOTOQs studied here, the background quasar is not bright enough in UV to measure $N$(\HI). \citet{kulkarni2022} have measured $N$({\HI}) towards a handful of GOTOQs. These sources do not show a strong trend between $N$({\HI}) and $E(B-V)$. It is not clear whether the lack of correlation is because of selection bias where one tends to pick less reddened lines-of-sight for UV spectroscopic observations. On the other hand, based on stacking of SDSS spectra in \MgII\ absorption-selected samples, quasars with \HI\ 21-cm absorption detected towards them have been found to be redder on average \citep{Dutta2017FeII}.

\vskip 0.1 in
\noindent{\bf Future Perspective:}
In the near future, data releases from wide-field, fiber-based spectroscopic surveys such as DESI \citep{desi_collab2016}, the 4-Metre Multi-Object Spectrograph Telescope \citep[4MOST;][]{deJong2019}, and the WHT Enhanced Area Velocity Explorer \citep[WEAVE;][]{Jin2024} are expected to significantly expand the catalog of known GOTOQs. Notably, DESI alone is set to provide approximately 4 million optical quasar spectra -- an increase of five times the number of quasars with optical spectra compared to SDSS. In the early data release from DESI, covering a mere 2\% of the DESI sky and featuring around 90,000 quasar spectra, we have identified about 200 GOTOQs at $z \lesssim 1$ (Guha et al, in prep). Consequently, it is expected that the final sample will comprise of around 10000 GOTOQs. Assuming a 1\% incidence of radio-loud quasars ($f_{peak} \geqslant 30$ mJy) among these as found in our parent sample of GOTOQs, we anticipate about 100 GOTOQs with radio-loud quasars in the background. We plan to carry out a systematic survey of \HI\ 21-cm and OH 18-cm absorption from these GOTOQs using uGMRT, VLA, and MeerKAT (for the southern hemisphere). This will substantially increase our ability to statistically map the cold neutral and molecular gas distribution around galaxies and study their redshift evolution, given the success of our pilot survey.

\section{acknowledgments}
This work makes use of the following softwares and packages: \textsc{casa} \citep{McMullin2007}, \textsc{aoflagger} \citep{Offringa2012}, \textsc{numpy} \citep{numpy2020}, \textsc{matplotlib} \citep{matplotlib2007}, \textsc{scipy} \citep{scipy2020}, \textsc{astropy} \citep{astropy:2018}, \textsc{dust\_extinction} \citep{Gordon2024}, and \textsc{ultranest} \citep{Buchner2021}.



%\appendix

\bibliography{sample631}{}
\bibliographystyle{aasjournal}



\appendix
\section{Radio morphology of the background quasars}
Here, we show the radio contours obtained from the uGMRT 1.4 GHz observations overlaid on the DESI r-band images centered on background quasars of the GOTOQs. The centroid of the radio contours are well-aligned with the optical centroids of the background quasars.

\begin{figure*}[htbp]
    \centering
    % First figure
    \begin{minipage}[b]{0.32\textwidth}
        \centering
        \includegraphics[width=\textwidth]{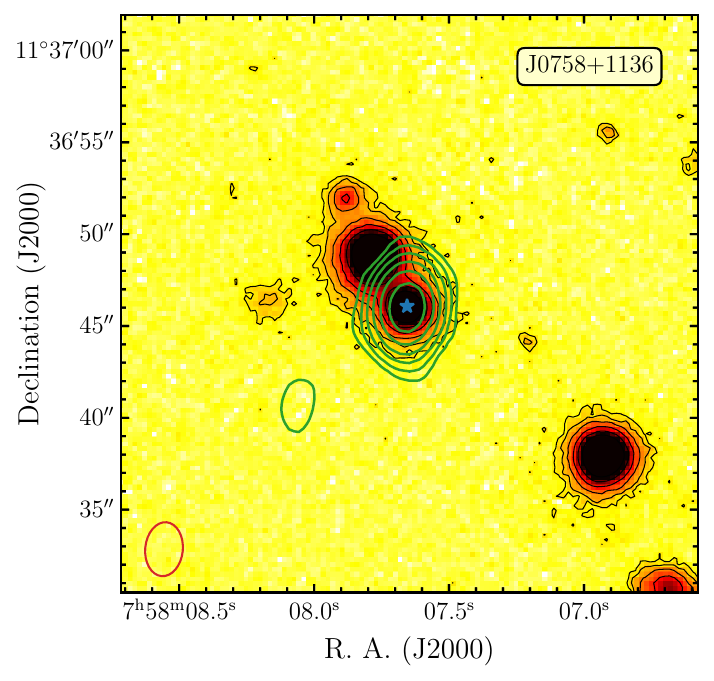}
    \end{minipage}
    \hfill
    % Second figure
    \begin{minipage}[b]{0.32\textwidth}
        \centering
        \includegraphics[width=\textwidth]{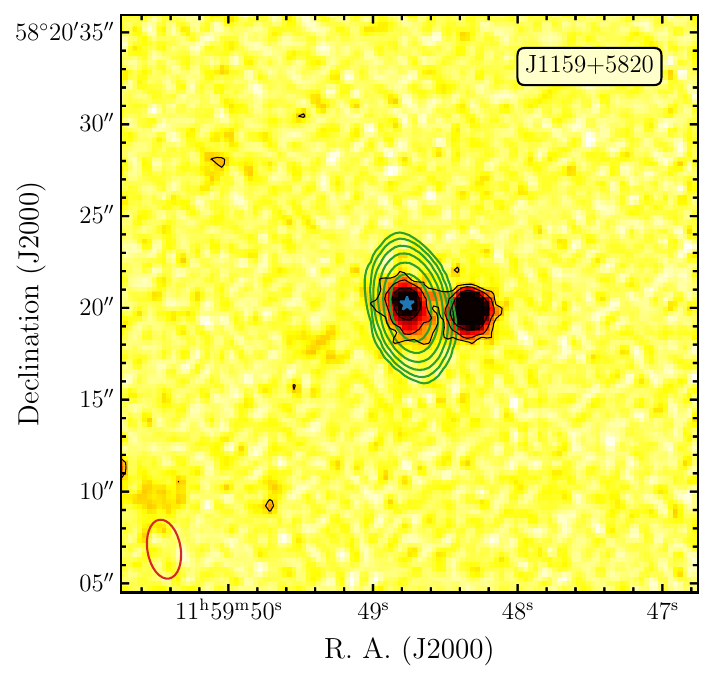}
    \end{minipage}
    \hfill
    % Third figure
    \begin{minipage}[b]{0.32\textwidth}
        \centering
        \includegraphics[width=\textwidth]{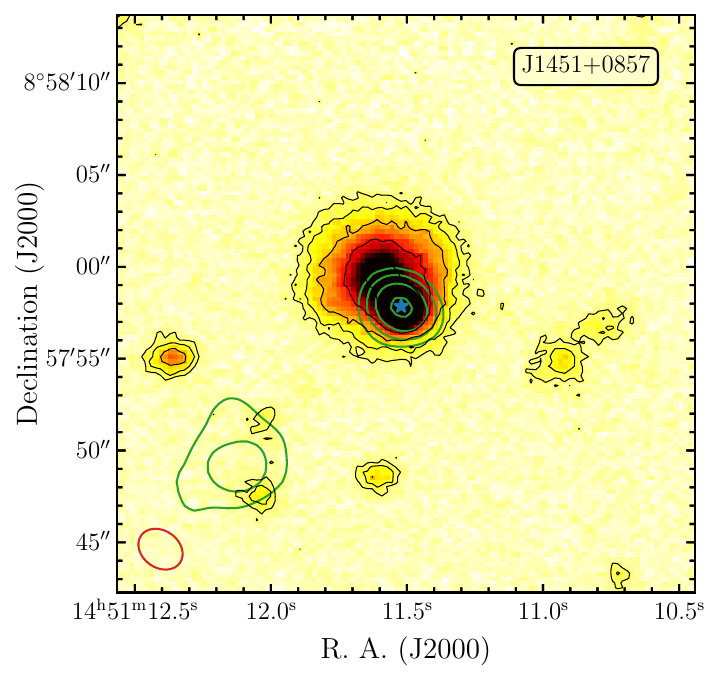}
    \end{minipage}
    \caption{DESI-LIS $r$-band continuum images of the GOTOQs observed with the uGMRT. The green contours represent the radio continuum flux density from uGMRT observations with a 6.25 MHz bandwidth, centered on the expected \HI\ 21-cm absorption frequency. The outermost green contour in each panel corresponds to $10\sigma$ above the median noise level, with successive inner contours increasing by factors of 2. The blue star marks the optical centroid of the background quasar. The black contours in each panel corresponds to $3\sigma$ above the median noise level, with successive inner contours increasing by factors of 2. The uGMRT synthesized beam is shown with the red ellipse at the bottom-left corner in each panel. 
    %{\bf All our radio sources are unresolved in our uGMRT images.}
    }
    \label{fig:radio_contours}
\end{figure*}

\section{Line of sight reddening of the background quasars by the GOTOQs}
\label{sec:appendix_reddening}
Since the absorber rest-frame wavelengths do not extend below 3000\AA, the choice of extinction curve has little impact on the results. We obtain similar values of reduced $\chi^2$ and $E(B-V)$ when using SMC, LMC, or Milky Way extinction curves. Therefore, in the appendix of the main paper, we present the dust extinction fits for the GOTOQs studied in this work assuming an SMC-like extinction curve.

\begin{figure}
    \centering
    \includegraphics[width=0.81\linewidth]{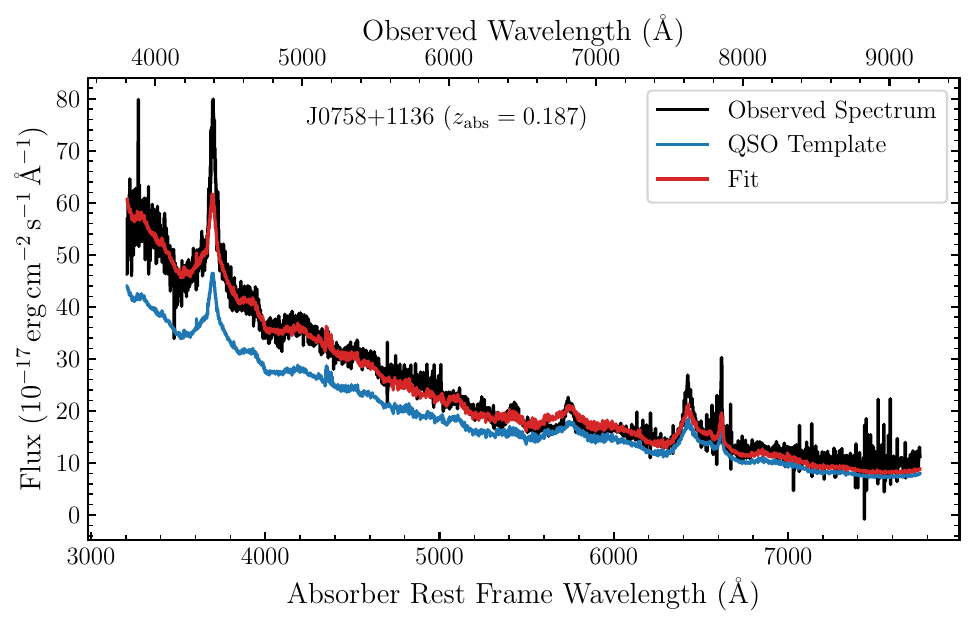}
    \caption{Estimation of the line-of-sight reddening of the background quasar caused by the foreground galaxy for the GOTOQ J0758+1136. The black spectrum is the observed spectrum of the background quasar, while the blue spectrum shows the template quasar spectrum redshifted to the redshift of the quasar. The red curve corresponds to the best-fit spectrum, obtained by applying a color excess of $E(B-V) = -0.071$ to the template spectrum at the redshift of the foreground galaxy. In this case the quasar spectrum tend to be bluer than the template spectrum used. Such measurements are useful for quantifying the scatter in E(B-V) propagating from spread in the QSO SEDs.}
    \label{fig:J0758_reddening}
\end{figure}
\begin{figure}
    \centering
    \includegraphics[width=0.81\linewidth]{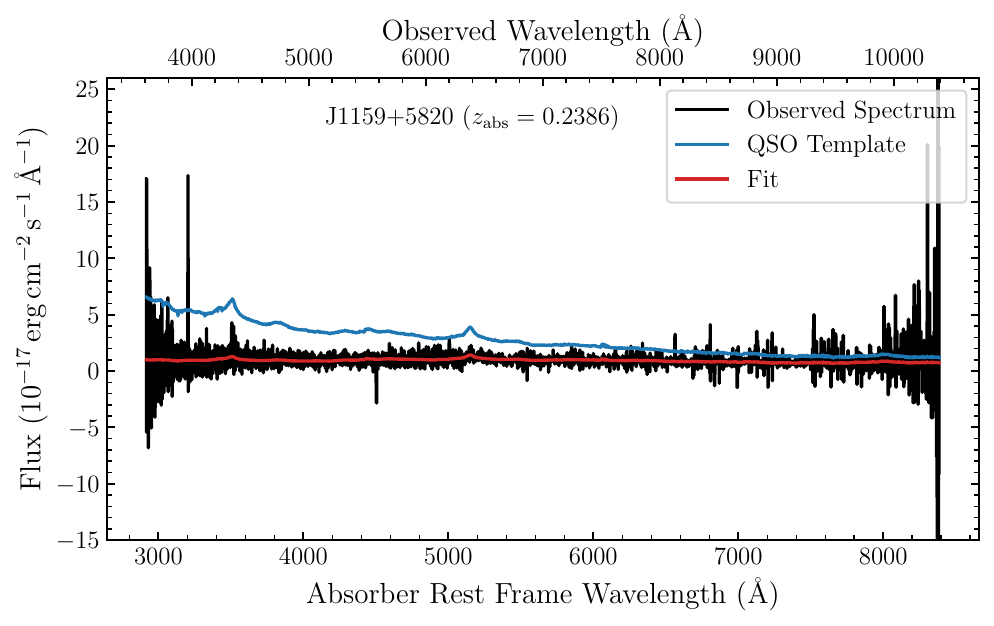}
    \caption{Same as the above figure, but for the GOTOQ J1159+5820.}
    \label{fig:J1159_reddening}
\end{figure}
\begin{figure}
    \centering
    \includegraphics[width=0.81\linewidth]{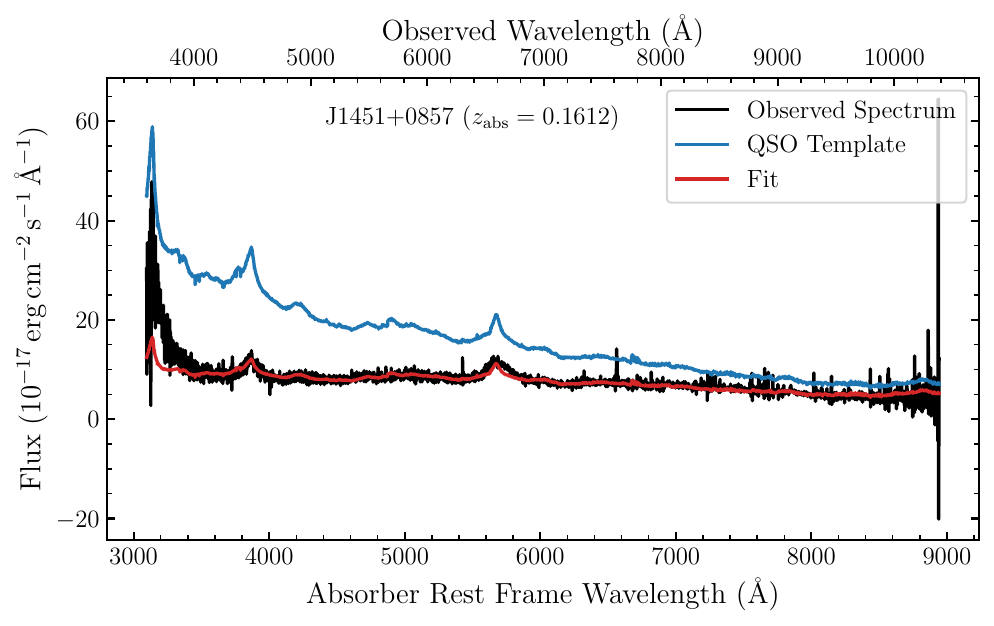}
    \caption{Same as the above figure, but for the GOTOQ J1451+0857.}
    \label{fig:J1451_reddening}
\end{figure}

\end{document}